\documentclass[%
 aip,
 amsmath,amssymb,
 reprint,%
]{revtex4-2}

\usepackage{graphicx}
\usepackage{dcolumn}
\usepackage{bm}

\usepackage[utf8]{inputenc}
\usepackage[T1]{fontenc}
\usepackage{mathptmx}
\usepackage{etoolbox}
\usepackage{xcolor}

\makeatletter
\def\@email#1#2{%
 \endgroup
 \patchcmd{\titleblock@produce}
  {\frontmatter@RRAPformat}
  {\frontmatter@RRAPformat{\produce@RRAP{*#1\href{mailto:#2}{#2}}}\frontmatter@RRAPformat}
  {}{}
}%
\makeatother
\begin{document}

\preprint{AIP/123-QED}

\title{Free-space coupling and characterization of transverse bulk phonon modes in lithium niobate in a quantum acoustic device}

\author{J.~M.~Kitzman}
\affiliation{Department~of~Physics~and~Astronomy,~Michigan~State~University,~East~Lansing,~Michigan~48824,~USA}
\author{J.~R.~Lane}
\affiliation{Department~of~Physics~and~Astronomy,~Michigan~State~University,~East~Lansing,~Michigan~48824,~USA}
\author{C.~Undershute}
\affiliation{Department~of~Physics~and~Astronomy,~Michigan~State~University,~East~Lansing,~Michigan~48824,~USA}
\author{M.~Drimmer}
\affiliation{Department of Physics, Eidgen$\ddot{o}$ssiche Technische Hochschule Z$\ddot{u}$rich, 8093 Z$\ddot{u}$rich, Switzerland}
\affiliation{Quantum Center, Eidgen$\ddot{o}$ssiche Technische Hochschule Z$\ddot{u}$rich, 8093 Z$\ddot{u}$rich, Switzerland}
\author{A.~J.~Schleusner}
\affiliation{Department~of~Physics~and~Astronomy,~Michigan~State~University,~East~Lansing,~Michigan~48824,~USA}
\author{N.~R.~Beysengulov}
\affiliation{Department~of~Physics~and~Astronomy,~Michigan~State~University,~East~Lansing,~Michigan~48824,~USA}
\author{C.~A.~Mikolas}
\affiliation{Department~of~Physics~and~Astronomy,~Michigan~State~University,~East~Lansing,~Michigan~48824,~USA}
\author{J.~Pollanen$^{*}$}
\email[]{Author to whom correspondence should be addressed: pollanen@msu.edu}
\affiliation{Department~of~Physics~and~Astronomy,~Michigan~State~University,~East~Lansing,~Michigan~48824,~USA}

\date{\today}
\begin{abstract}
Transverse bulk phonons in a multimode integrated quantum acoustic device are excited and characterized via their free-space coupling to a three-dimensional (3D) microwave cavity. These bulk acoustic modes are defined by the geometry of the $Y$-cut lithium niobate substrate in which they reside and couple to the cavity electric field via a large dipole antenna, with an interaction strength on the order of the cavity line-width. Using finite element modeling (FEM) we determine that the bulk phonons excited by the cavity field have a transverse polarization with a shear velocity matching previously reported values. We demonstrate how the coupling between these transverse acoustic modes and the electric field of the 3D cavity depends on the relative orientation of the device dipole, with a coupling persisting to room temperature. Our study demonstrates the versatility of 3D microwave cavities for mediating contact-less coupling to quantum, and classical, piezoacoustic devices.
\end{abstract}
\maketitle
\noindent Quantum acoustics systems, in which superconducting qubits are interfaced with mechanical degrees of freedom offer a promising platform for quantum information science, as mechanical resonators having a small spatial footprint and long coherence times can be straightforwardly fabricated. By engineering hybrid quantum systems in this fashion, it is possible to design quantum memory protocols~\cite{Hann2019}, implement microwave-to-optical transduction schemes~\cite{Mirhosseini2020,Delaney2022}, and operate the qubit as a sensor~\cite{AYCleland2023,Kitzman2023}. The ability to leverage mechanical degrees of freedom at the level of single, or few phonons, makes them promising candidates for quantum information processing, with recent experimental results demonstrating the creation of phononic Schr$\ddot{\textrm{o}}$dinger cat states~\cite{Bild2023}, the joint entanglement of high-frequency mechanical oscillators~\cite{Wollack2022}, and the ability to simulate open quantum acoustic systems~\cite{Kitzman2023_be}. However, the relatively large and geometrically complex structures typically used in quantum acoustics devices can host a litany of other mechanical modes that may have spurious coupling to the electromagnetic degrees of freedom of the quantum circuit. Identifying and understanding these couplings is vital for high-fidelity control of the device degrees of freedom, but also opens the door to coupling simultaneously to multiple distinct types of phononic excitations in a single architecture. On the other hand, when they are poorly controlled, these modes can act as unwanted decoherence channels for the quantum systems of interest. Optimization of device geometry allows for the mitigation of these unwanted couplings and methods for improving the coherence of quantum acoustic systems. In this work, we demonstrate a hybrid system that enables the interaction between a microwave cavity and natural \emph{bulk transverse} substrate phonons in \textit{Y}-cut lithium niobate. By integrating this hybrid electromechanical system with a superconducting qubit, we are able to tune the cavity frequency via its dispersive interaction with the qubit, enabling measurement of the coupling between the cavity and high-frequency overtones of the acoustic mode. These transverse acoustic modes are directly excited via the free-space coupling of the device dipole to the 3D microwave cavity housing the devices and at low temperature we find that they exhibit coupling to the qubit-cavity system with an interaction strength comparable to the decay rate of the hybrid device. Additionally we find that similar devices, in which the qubit is removed, enable contact-less coupling to these naturally occurring transverse modes with a coupling that persists to room temperature~(cf.~Ref.~\cite{Zhai2022}).

The experimental results reported here were obtained using a flip-chip hybrid quantum acoustic device originally designed to investigate the coupling of a flux tunable transmon qubit, fabricated on silicon, to a surface acoustic wave (SAW) resonator, fabricated on \textit{YZ}-lithium niobate~\cite{Kitzman2023_be,Kitzman2023}. These previous experiments reached the strong coupling regime of quantum acoustics, with a mechanical coupling strength of $g/(2\pi)~\approx~10$~MHz between the qubit and SAW modes. In the present manuscript, we report on experimental work demonstrating the existence of additional \textit{bulk} phononic modes in the lithium niobate chip and how their direct coupling to the resonant mode of a 3D microwave cavity is mediated by the dipole moment of the device. Due to the dispersive interaction between the cavity and the qubit~\cite{Blais2021}, these bulk acoustic modes also indirectly interact with the qubit. As shown in Fig.~1~(a), the qubit and phononic chips are coupled purely capacitively to each other via large (250 $\mu$m)$^2$ antenna pads on either substrate that form a set of parallel plate capacitors when the devices are vertically aligned and adhered together using standard flip-chip techniques~\cite{sat19}.
\begin{figure*}[ht!]
    \centering
    \includegraphics[width = 16cm]{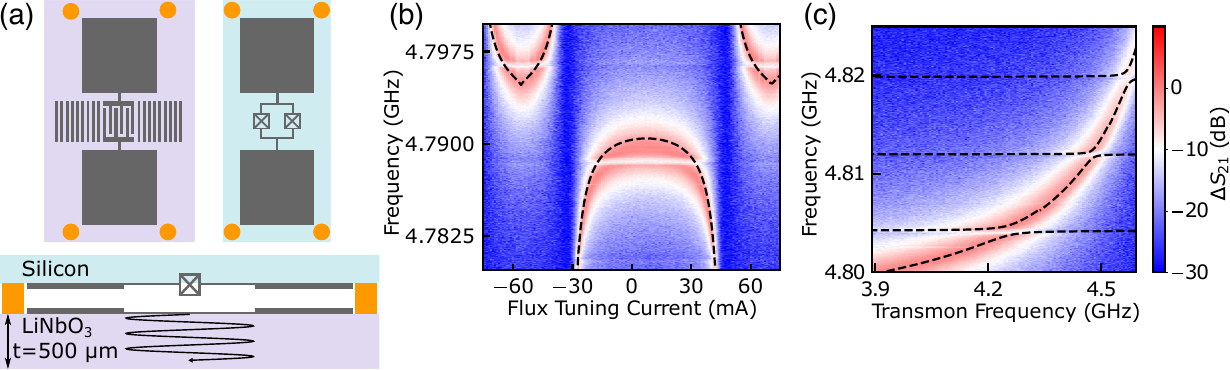}
    \caption{(a) Schematic of the quantum acoustic device (not to scale). Top: Top view of the lithium niobate chip and the qubit chip before flip-chip assembly. Bottom: Side view of the composite device post flip-chip assembly. The transverse bulk modes excited in the lithium niobate are schematically represented by the wavy line. The transmon qubit is fabricated by depositing aluminium (grey) on silicon (blue), similarly the acoustic resonator is fabricated by depositing aluminium on lithium niobate (purple). Patterned circles of photoresist (orange) are used as a spacer between chips. Once the composite device is flip-chip assembled it is placed in the center of a 3D microwave cavity to probe the properties of the quantum acoustic system. (b) Spectroscopy of the hybrid qubit-cavity system as the qubit frequency is tuned via the magnetic flux through an external superconducting coil. The qubit and cavity are coupled with an interaction strength of $g/(2\pi) = 73~\pm~2$~MHz (black dashed line). Additional coupling between the hybrid system and bulk acoustic modes is observed as horizontal features at constant frequency, which are largely independent of the external magnetic flux. (c) Spectroscopic data over a tighter frequency range demonstrating clear avoided crossings between the qubit-cavity system and the acoustic modes. The fit to the data (dashed black lines) represent a coupled oscillator model used to extract the coupling strengths (see~Fig.~\ref{f2}).}
    \label{f1}
\end{figure*}
The flip-chip stack is mounted in a copper 3D electromagnetic cavity to control and measure the coupled devices (see Fig.~1~(a)). The antenna pads on the qubit and phononic chips serve dual purpose, enabling the coupling between the devices as well as serving as a coupling mechanism between the electric field in the microwave cavity and either device independently~\cite{Kitzman2023}. As we will describe below, these macroscopic antenna pads also introduce an unintended and surprisingly strong coupling to high-overtone bulk phononic excitations, with harmonic number beyond 1000, in the lithium niobate substrate. 

\begin{figure}[b]
    \centering
    \includegraphics[width = 7cm]{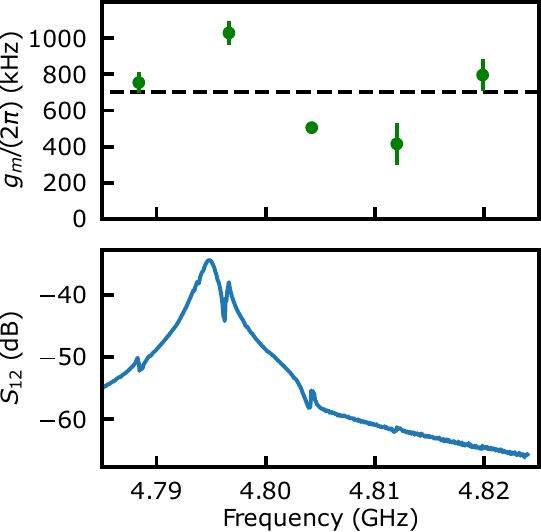}
    \caption{Top: Coupling strength, $g_{m}/(2\pi)$ between the microwave cavity and five overtone bulk mechanical modes. The black dashed line represents the average coupling strength for these five modes. Bottom: Measurement of the cavity spectrum at large input power ($P_{\textrm{input}} \approx -60~\textrm{dBm}$, corresponding to $\approx 10^8$ cavity photons~\cite{aspelmeyer2014cavityo}) demonstrating that the transmission is independent of the qubit state as described in the main text. Coupling to the acoustic modes manifests as characteristic ``notches" in the cavity transmission.}
    \label{f2}
\end{figure}

To experimentally investigate the coupling of the microwave cavity to these modes we measure the microwave transmission $S_{12}$ through the 3D microwave cavity at a relatively low input power ($P_{\textrm{input}}\approx-110~\textrm{dBm}$) as we tune the qubit frequency via an external magnetic flux. As shown in Fig.~\ref{f1}(b), when the resonant frequency of the qubit is tuned across an entire flux quantum, the qubit and cavity modes undergo an avoided crossing, which can be used to extract a coupling strength of $g/(2\pi) = 73\pm2$~MHz between the two modes. As seen in Fig. 1(b) the cavity transmission also exhibits a series of avoided crossings, indicating coupling to additional modes, which appear as horizontal breaks in microwave transmission, and are largely independent of external magnetic flux. By zooming in on several of these additional avoided crossings in Fig.~\ref{f1}~(c) the transmission data are fit to a coupled oscillator model to extract the interaction strength between the qubit-cavity mode and these additional weaker resonances. In the top panel of Fig.~\ref{f2} we present the fitted coupling rates $g_{m}/(2\pi)$ for the five anti-crossings that are within a few line-widths of the resonant frequency of the microwave cavity. The bottom panel of Fig.~\ref{f2} shows how the coupling to these modes manifests in the bare cavity transmission when the cavity is flooded with a large number of photons ($P_{\textrm{input}} \approx -60~\textrm{dBm}$)~\cite{Reed2010}. Even when the cavity is populated with many photons, and the qubit state no longer dispersively shifts the cavity resonance, the presence of these additional modes is still strongly imparted on the cavity spectrum as ``notches" in the measured transmission. 

Additional acoustic modes in the strongly piezoelectric lithium niobate substrate are a natural explanation for these additional resonances, which couple to the hybrid qubit-cavity system. In fact, periodically notched spectra, similar to the one shown in the bottom panel of Fig.~\ref{f2}, have recently been observed in the transmission of microwave cavities coupled to high-overtone longitudinal phononic modes in quartz~\cite{Yoon23}. Additionally, high-overtone bulk acoustic resonators (HBARs), using a variety of piezoelectric materials, have been shown to strongly couple to superconducting qubits in quantum acoustics architectures~\cite{chu17,chu18,Kervinen2018,Kervinen2019,vonLupke2022,vonLupke2023,Bild2023,Crump2023}. Furthermore, as we describe below, we use a combination of finite element modeling (FEM) and additional characterization measurements to verify the nature of these bulk acoustic modes, their polarization, as well as their coupling mechanism to the hybrid qubit-cavity system.

\begin{figure}[t]
    \centering
    \includegraphics[width = 5.5cm]{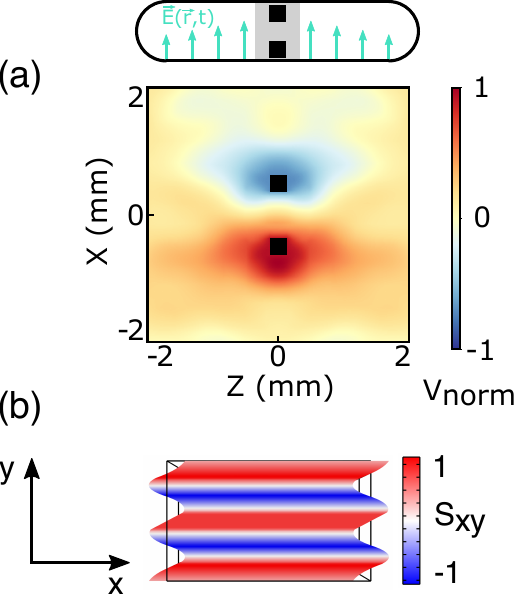}
    \caption{a) Top: Schematic indicating the orientation of the time-varying electric field of the 3D microwave cavity. Bottom: COMSOL simulation of the induced normalized voltage on the surface of the lithium niobate substrate when the system is driven at the fundamental frequency of the transverse mode $\omega/(2\pi) = 3.5784$~MHz. The black squares represent the antenna pads, and the profile of the induced voltage has a dipole spatial structure. (b) FEM simulation of the transverse strain, $S_{\textrm{xy}}$, associated with the $5^{th}$ harmonic of the transverse bulk acoustic mode.}
    \label{f3}
\end{figure}

As shown in the COMSOL Multiphysics simulations presented in Fig.~\ref{f3}(a), the time-varying electric field within the microwave cavity will impinge on the large-scale aluminum antenna pads on the surface of the lithium niobate chip. These two pads function as an electric dipole and the mobile charges in these metallic pads screen the cavity electric field, locally polarizing the substrate. As a consequence of the piezoelectric effect, this polarization creates strain within the lithium niobate substrate. If the frequency of the time-varying electromagnetic cavity field is resonant with a natural mechanical mode imposed by the geometry of the substrate and the speed of sound in lithium niobate, the generated strain will excite acoustic oscillations that will resonate within the bulk of the material. In this fashion, the dipole antenna on the lithium niobate chip enables the interaction between the cavity mode and bulk phonon modes in the substrate. Furthermore, since the resonant frequency of the cavity can be weakly flux-tuned via its dispersive interaction with the qubit, this device architecture allows us to measure the interaction between the cavity mode and acoustic modes over a frequency range larger than the cavity line-width.
To determine the exact nature of the acoustic modes being excited by this dipole-coupling to the cavity field, we use COMSOL simulations to find the mechanical eigenmodes of the \textit{Y}-cut lithium niobate substrate as well as their polarization~\cite{Bandrier2023}. These simulations reveal the existence of both longitudinal and transverse acoustic modes. In order to model the effect of the cavity electric field, we apply a time-varying voltage difference across the chip in the \textit{x}-direction of the substrate in these simulations (see Fig.~\ref{f3}(a)). In the presence of this external electric field, the symmetry of the combined electromechanical system only supports mechanical modes that mirror the symmetry of the external electric field, and thus we find that only transversely polarized mechanical modes can be excited (see Fig.~\ref{f3}(b) for the spatial structure of the $n = 5$ mode). In particular, the results of the simulations reveal a transverse mode having fundamental frequency $f_1 = 3.5784~$MHz propagating along the \textit{y}-axis of the substrate. We note that the simulations enforce that the mechanical displacement field is equal along the two edges of the substrate parallel to the $x$-axis. The fundamental frequency of the transverse mode, and its harmonics, are related to the speed of bulk transverse sound in lithium niobate $v_t$ and the thickness of the chip:
\begin{equation}\label{modes}
    f_n = \frac{nv_t}{2t},
\end{equation}
where $n$ indexes the harmonic number and $t = 500~\mu$m is the thickness of the substrate used in our experiments and simulations. The speed of transverse sound within the bulk lithium niobate is related to the material properties of the substrate via Hooke's Law:
\begin{equation}\label{elastic}
    C_{44} = \rho v_t^2,
\end{equation}
where $C_{44} = 5.95\times10^{10}$~Pa is the appropriate elastic constant for transverse mechanical waves in \textit{Y}-cut lithium niobate and $\rho = 4647$~kg/m$^3$ is the mass density of the material~\cite{Ahmadi2004}. Eqn.~\ref{elastic} gives the speed of transversely polarized bulk sound as 3578~m/s, consistent with Ref.~\cite{Laude2005}, and a corresponding free spectral range of $\Delta f_n= 3.578$~MHz between adjacent acoustic modes.

When a spatially uniform external electric field ($E_{\textrm{ext}}$) is applied across the substrate, as is the case for the cavity field in the vicinity of the dipole antenna on the lithium niobate chip in our experiments, the piezoelectric coupling $g_{\textrm{em}}$ is proportional to the overlap integral between the electric and strain fields~\cite{Yoon23}:
\begin{equation}\label{interaction}
g_{\textrm{em}} \propto \int_{V} E_{\textrm{ext}}~S(\vec{r})~ \textrm{d}V,
\end{equation}
where $S(\vec{r})$ is the spatially varying strain field within the chip and the integral is taken over the entire volume $V$. The $y$-dependence of the strain field for transverse phonons is given by $S(y) = S_0\sin(\frac{2 \pi y}{\lambda_n})$, where $S_0$ is the amplitude of the strain. For the $n^{\textrm{th}}$ harmonic of the transverse mode having wavelength $\lambda_n = 2t/n$, Eqn.~\ref{interaction} predicts non-zero coupling for odd-indexed modes only. This implies that in our experiments the 3D microwave cavity electric field should couple only to every other transverse phonon mode, leading to an effective free spectral range $2\times f_n \simeq 7.16~$MHz based on the fundamental frequency determined from the FEM simulations. This matches (to within 10\%) the measured free spectral range determined from the spacing of adjacent notches in the microwave cavity transmission shown in Fig.~\ref{f2}. Based on the calculated free spectral range of the bulk acoustic modes and the cavity frequency, we can reliably excite bulk mechanical excitations with mode number $n \simeq 1339$ that couple to the qubit.
\begin{figure}[t]
    \centering
    \includegraphics[width = 7.5cm]{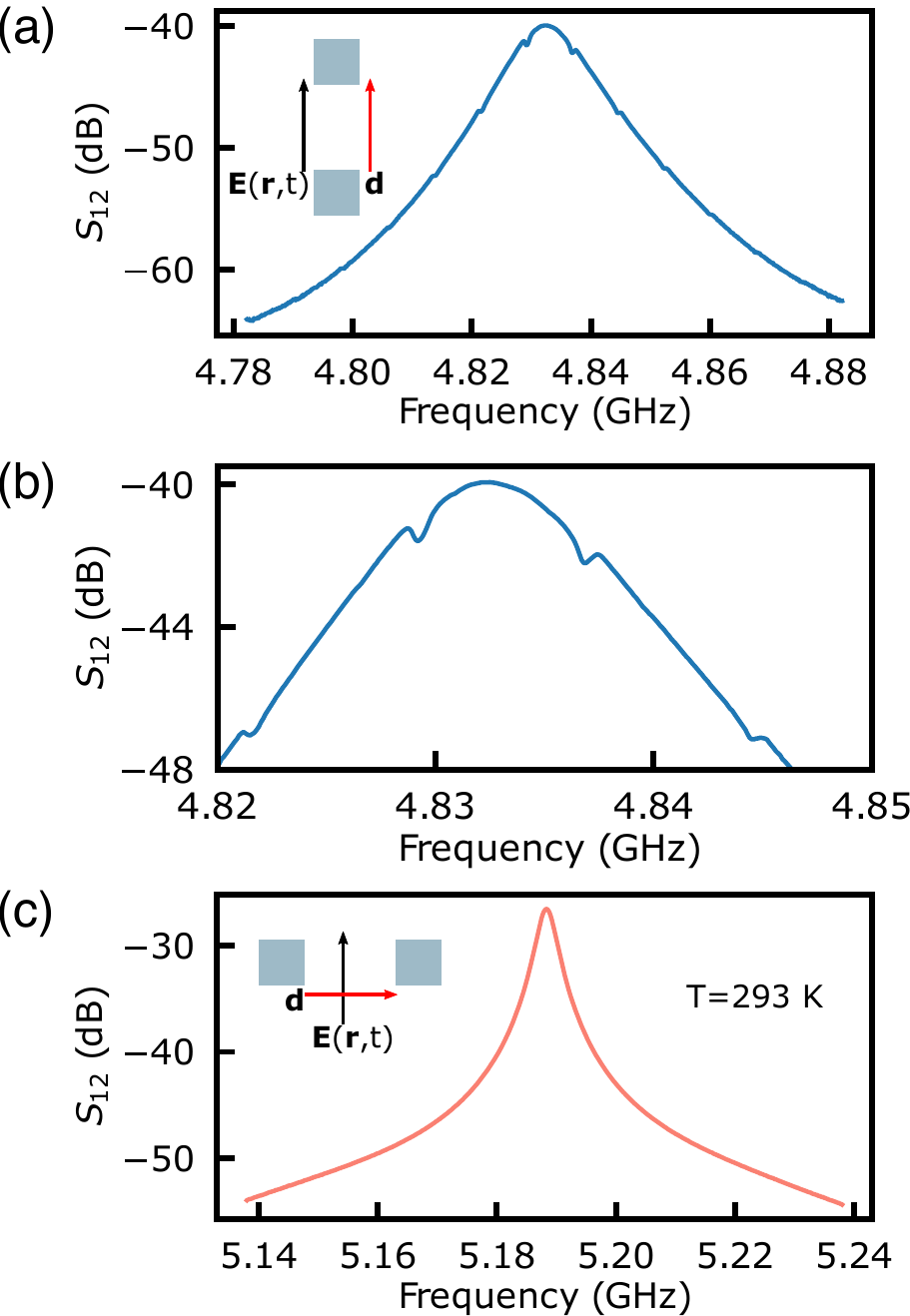}
    \caption{Room temperature characterization of the transverse acoustic modes. (a) When the dipole moment of the metallic pads ($\vec{d}$) is parallel to the cavity electric field, coupling is observed between the cavity mode and the acoustic modes of the substrate. (b) Zoomed in cavity transmission from (a) near the cavity resonance. The coupling to the bulk acoustic modes is apparent as ``notches" in the cavity transmission. (c) When the LiNbO$_{3}$ substrate is oriented such that the dipole moment is perpendicular to the cavity field, no coupling between the two systems is observed. The shift in the cavity frequency between panels (a,b) and (c) arises from the large anisotropy of the lithium niobate dielectric tensor~\cite{Ahmadi2004}.}
    \label{f4}
\end{figure}

In addition to spectroscopy measurements at cryogenic temperatures, we also performed additional characterization experiments at room temperature to investigate these transverse bulk acoustic modes, and their coupling to the microwave cavity. In particular, to further elucidate the coupling of the cavity field to the dipole formed by the metallic antenna pads, we fabricated an additional device having only large antenna pads on a separate $Y$-cut lithium niobate chip and loaded it into a second 3D microwave cavity. As shown in Fig.~\ref{f4}(a,b) the appearance of the notches in the cavity transmission clearly persists even at room temperature, albeit with a reduced amplitude. These measurements serve to reinforce the conclusion that the 3D cavity enables a method for contact-less excitation of these high-overtone acoustic modes. Additionally, we find that when the sample is rotated by $90^\circ$ in the cavity, the notches in the microwave transmission disappear indicating that the coupling to the transverse modes vanishes (see Fig.~\ref{f4}(c)). In this configuration the dipole moment of the device is perpendicular to the cavity electric field and cannot induce a polarization between the pads, nor the reciprocally generated transverse strain field via the piezoelectricity of the substrate. This rotation dependence is consistent with the conclusions we draw from the FEM simulations of the device and further verify that the dipole coupling between the pads and the cavity field is responsible for exciting the transverse modes of the lithium niobate chip. Finally, comparing Fig.~\ref{f4}(a) with Fig.~\ref{f4}(c) clearly shows that the microwave cavity line-width is significantly broadened when the lithium niobate chip is oriented such that the antenna pads are aligned to the cavity field. In particular, we find that in this configuration the cavity quality factor is decreased by a factor of approximately $3.75$, from $Q \simeq 1500$ to $Q \simeq 400$, relative to the configuration in which the pads are orthogonal to the cavity field. To understand the origin of these additional losses, we tested bare lithium niobate chips having the same dimensions but without the metallic antenna pads. Unsurprisingly, in these additional control samples we do not observe any coupling between the cavity field and bulk acoustic modes regardless of the orientation of the chip in the microwave cavity (i.e. no cavity notches are observed). Interestingly, however, we find that the cavity quality factor is $Q \simeq 1500$ regardless of the chip orientation, strongly indicating that the reduction in the cavity quality factor observed in Fig.~\ref{f4}(a,b) arises from the mechanical loss associated with the transverse phonon modes. These findings imply that the transverse phonons modes have a relatively low quality factor at room temperature. Analysis of the linewidths of the modes in Fig.~\ref{f1} allows us to estimate the low temperature quality factor of the modes, which we find to be on the order of several thousand. These values are significantly lower than the quality factors of bulk modes previously reported in quantum acoustic experiments~\cite{Bild2023, chu17, chu18}. In these experiments the acoustic resonators were designed to host bulk phonon modes with high quality factors and well-defined spatial mode volume. In contrast, the original purpose of the device described here was to host surface acoustic wave modes and the device was not optimized to efficiently transduce nor confine bulk acoustic modes. In fact, in other experiments in which planar electromagnetic resonators have demonstrated spurious coupling to mechanical modes, these mechanical modes exhibit low quality factors (and weak coupling) and act as a loss channel for the electromagnetic resonator~\cite{scigliuzzo2020phononic}. As a final remark, we note that we observe a reduction in the microwave cavity quality factor by a factor of $~1.5$ when it contains an unpatterned lithium niobate chip at room temperature, which is consistent with reported values of this material's dielectric loss~\cite{yang2007characteristics, wong2002properties}.

In conclusion, we have demonstrated the capacitive dipole coupling between a hybrid 3D cavity system and high-overtone transverse bulk acoustic modes in \textit{Y}-cut lithium niobate. These transverse acoustic modes are excited by the microwave field of the 3D cavity and detected as absorption features in the cavity transmission. The coupling to the acoustic modes is found to be comparable to other losses in the hybrid system, reinforcing the importance of understanding the role of unintended mechanical modes in quantum acoustic systems proposed for quantum applications. Beyond quantum acoustics, we find that this contact-less method of acoustic excitation and detection in our devices persists to room temperature, highlighting the utility and versatility of 3D cavities for characterizing the frequency response and mechanical losses in piezoacoustic devices in general. 
\section*{Data Availability}
The data that support the findings of this study are available from the corresponding authors upon reasonable request.
\section*{Acknowledgements}
We thank P.K. Rath and J. Zhang for fruitful discussions. This work was supported by the National Science Foundation (NSF) via grant number ECCS-2142846 (CAREER) and the Cowen Family Endowment. We also thank R. Loloee and B.~Bi for technical assistance and use of the W. M. Keck Microfabrication Facility at MSU.

%

\end{document}